\documentclass{caps}
\usepackage{peter}
\usepackage{graphicx}
\usepackage{epsf}

\begin{document}

\pagenumbering{arabic}


\author[Lifan Wang]{Lifan Wang\\
Lawrence Berkeley National Laboratory
}

\chapter{Supernova Explosions: Lessons from Spectropolarimetry}

\begin{abstract}%
Supernova can be polarized by an asymmetry in the explosion process, an off-center
source of illumination, scattering in an envelope distorted by rotation or a 
binary companion, or scattering by the circumstellar dust. Careful polarimetry
should thus provide insights to the nature of supernovae. Spectropolarimetry is the
most powerful tool to study the 3-D geometry of supernovae. A deep understanding of
the 3-D geometry of SNe is critical in using them for calibrated distance indicators.
\end{abstract}

\section{Introduction}

Polarimetry of supernovae (SNe) reveals the intrinsic ejecta asymmetries
(Shapiro \& Sutherland 1982, McCall 1984, H\"oflich 1991, 1996). SN~1987A represent a 
breakthrough in this area, by providing the first detailed record of the spectropolarimetric
evolution (e.g. Mendez et al. 1988; Cropper et al. 1988). SN~1993J also provided a wealth of 
data (Trammell, Hines, \& Wheeler 1993; Tran et al. 1997). Most of the theoretical 
interpretations of the polarimetry data are based on oblate or prolate spheroid
geometries. A very different picture of SN polarization is discussed in Wang \& Wheeler (1996)
where time-dependent dust scattering is shown to be a potential mechanism. 
New attempts are made with more complicated geometrical structures (Kasen et al. 2003; 2004).

We started a systematic program of supernova spectropolarimetry in 1995 using the 2.1 meter
telescope of the McDonald Observatory which nearly doubled the number of SNe with polarimetry
measurements in the first year of the program. The early qualitative conclusion was that 
core-collapse SNe are in general polarized at about 1\% level and that Type Ia are much 
less polarized ($\le 0.3\%$) (Wang et al. 1996). We have also observed several SNe 
for their spectropolarimetry using the Image Grism Polarimeter (IGP) at the 2.1 meter telescope 
and a polarimeter at 2.7 meter telescope. 
The spectropolarimetry data show that normal SN Ia are not highly polarized, but
a hint of intrinsic polarization is detected at levels around 0.3\% for SN~1996X at around 
optical maximum. We found also that a peculiar, subluminous SN~Ia 1999by was clearly polarized at 
a much higher level ($\sim$ 0.7\%, Howell et al. 2001) than normal SN~Ia. 
The polarization of core-collapse SNe evolve with time 
(Wang et al. 2001; Leonard et al. 2001), with the general trend being that the
polarization is larger post optical maximum than before or around optical maximum. Our 
observations indicate further that for core-collapse SNe the degree of polarization is 
larger for those with less massive envelopes (Wang et al 2001).
 
The ESO-VLT, with its extremely flexible ToO capability, makes it possible to study in great detail 
SN polarimetry. Not only can the SNe be follow to late nebular stages, but also they can be
observed early enough so that the most energetic outer layers can be probed. 
We have obtained high quality multi-epoch spectropolarimetry data of a normal Type Ia SN~2001el. 
The data allow us to probe both the mechanisms of SN Ia explosions, and the effect of asymmetry
on the use of SN Ia as distance indicators (Wang et al. 2003a).

For core-collapse events, the inferred degree of asymmetry is in general of the order of 10-30\%
if modeled in terms of oblate/prolate spheroids. The asphericity of the photosphere of SN Ia is 
perhaps not larger than 10-15\%. 

SN polarimetry is still a rapidly growing field in terms of both observations and
theories. Spectropolarimetry of SNe has been a driving force for 3-D models of SN Ia and
in turn the advance of 3-D models of SN explosion will help us to understand better the
observed phenomena. Here I will concentrate on the observational efforts of our SN
polarimetry program. 

\section{Core-collapse SNe}

\begin{figure}
\hspace{.0cm}
\includegraphics[width=12.cm,angle=0]{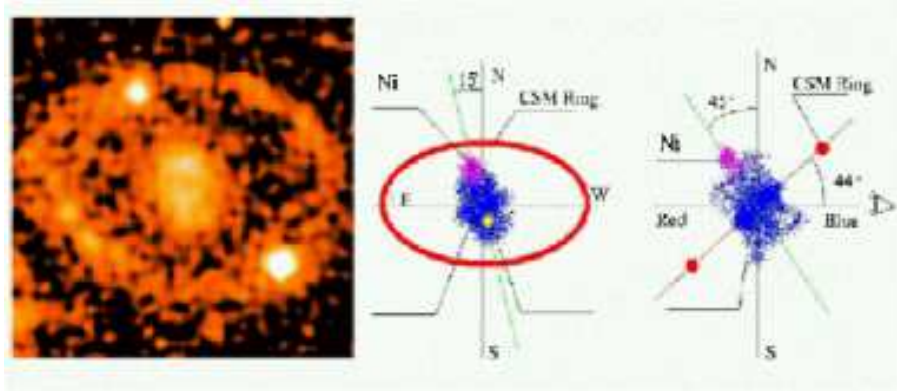}
\vspace{0cm}
\caption{Left: HST image of SN~1987A taken on 2000 June 11 with filter F439W.
The outer ellipse shows the CSM matter around the SN. The elongated central 
patch shows the SN~1987A ejecta. The apparent asymmetry of the SN ejecta 
represents the distribution of $^{56}$Fe. Middle and right: Schematic drawing 
of the inner CSM ring -- ejecta structure projected onto the plane of the 
sky (middle), and the plane of the line of sight and the direction to the 
North (right). The  symmetry axis of the
ejecta is about 14$^{\rm o}$ off to the east of the symmetry axis of the ring.
The $^{56}$Ni clump can be consistently placed at about 45$^{rm o}$ from the
line of sight. Oxygen is predominantly distributed on the plane of the 
CSM ring. There may also be an enhancement of dust opacity on the plane of
the ring which makes the image of the ejecta to appear fainter at the center.}
\end{figure}

\subsection{SN 1987A} The best observed core-collapse SN is, of course SN~1987A. 
Extensive Spectropolarimetry were obtained soon after the SN explosion. 
The polarimetry data were recently revisited by Wang et al. (2002) and compared 
with recent HST images and HST+STIS spectroscopy (see Fig. 1.1). By combining the early 
spectropolarimetry observations with the HST data we found a prolate geometry 
for the radioactive material at the inner most core of the ejecta, and an 
oblate spheroid geometry for the oxygen shell and the hydrogen envelope. The entire 
structure is approximately axially symmetric with the symmetry axis 
close to but noticeably offset from the minor axis of the circumstellar rings. 
These and more recent images from HST show an emission minimum at the center of the SN~1987A 
ejecta, which could be explained by an enhanced dust distribution on the plane 
defined by the CSM ring. 

\subsection{SN 1996cb}
\begin{figure}
\includegraphics[width=9.cm,angle=270]{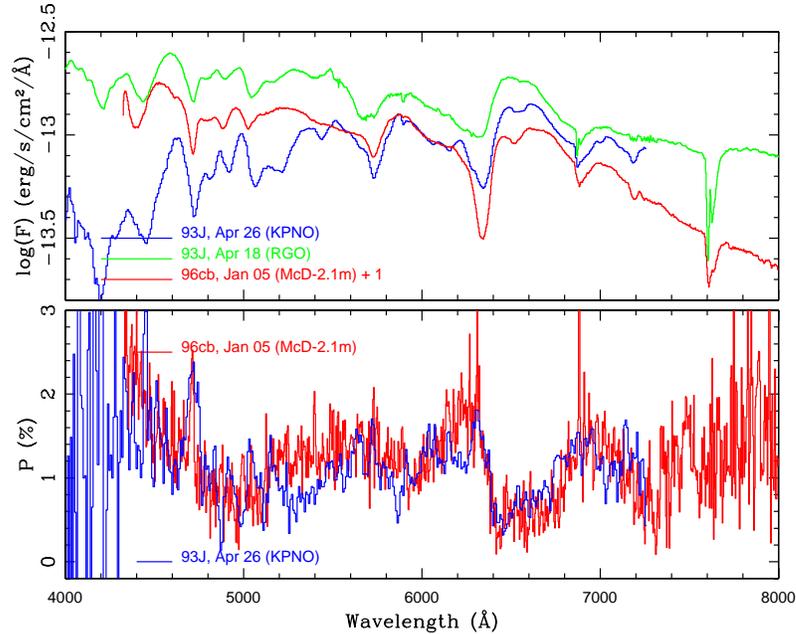}
\caption{Spectropolarimetry of SN~1996cb compared to those of 
SN~1993J obtained at comparable stages. The upper panel shows the spectroscopic data.
The lower panel shows the degree of polarization. The similarity of the 
polarization of the two SNe is striking.}
\end{figure}

SN 1996cb is a Type IIB SN whose photometric and spectroscopic behavior 
bear strong resemblances to that of SN 1993J (Fig. 1.2). These SNe show weak hydrogen 
features but strong helium lines. They are produced by 
stars which have lost most of their hydrogen envelope. Surprisingly, the 
spectropolarimetry data obtained at the McDonald Observatory show that they are 
also very similar to each other in spectropolarimetry. Detailed theoretical models were 
done for SN 1993J (H\"oflich et al. 1996). Based on those models, the SN 
photosphere can be highly aspherical with minor to major axis ratios around 0.6 
if modeled in terms of oblate spheroids. 

A natural question to ask is why SN~1993J and SN~1996cb are so similar despite 
the fact that they are so highly aspherical and clumpy (Wang, \& Hu 1994; Spyromilio 1994; 
Wang et al. 2001). The complementary question is whether these SNe are 
still Type IIB when viewed at different viewing angles. Our guess is that they 
are likely a select subgroup of a more common phenomena. More data, especially 
spectropolarimetry, will help to create a unified picture of SN IIB.

\subsection{SN 1997X}

\begin{figure}
\hspace{-1.15cm}
\includegraphics[width=5.cm,angle=0]{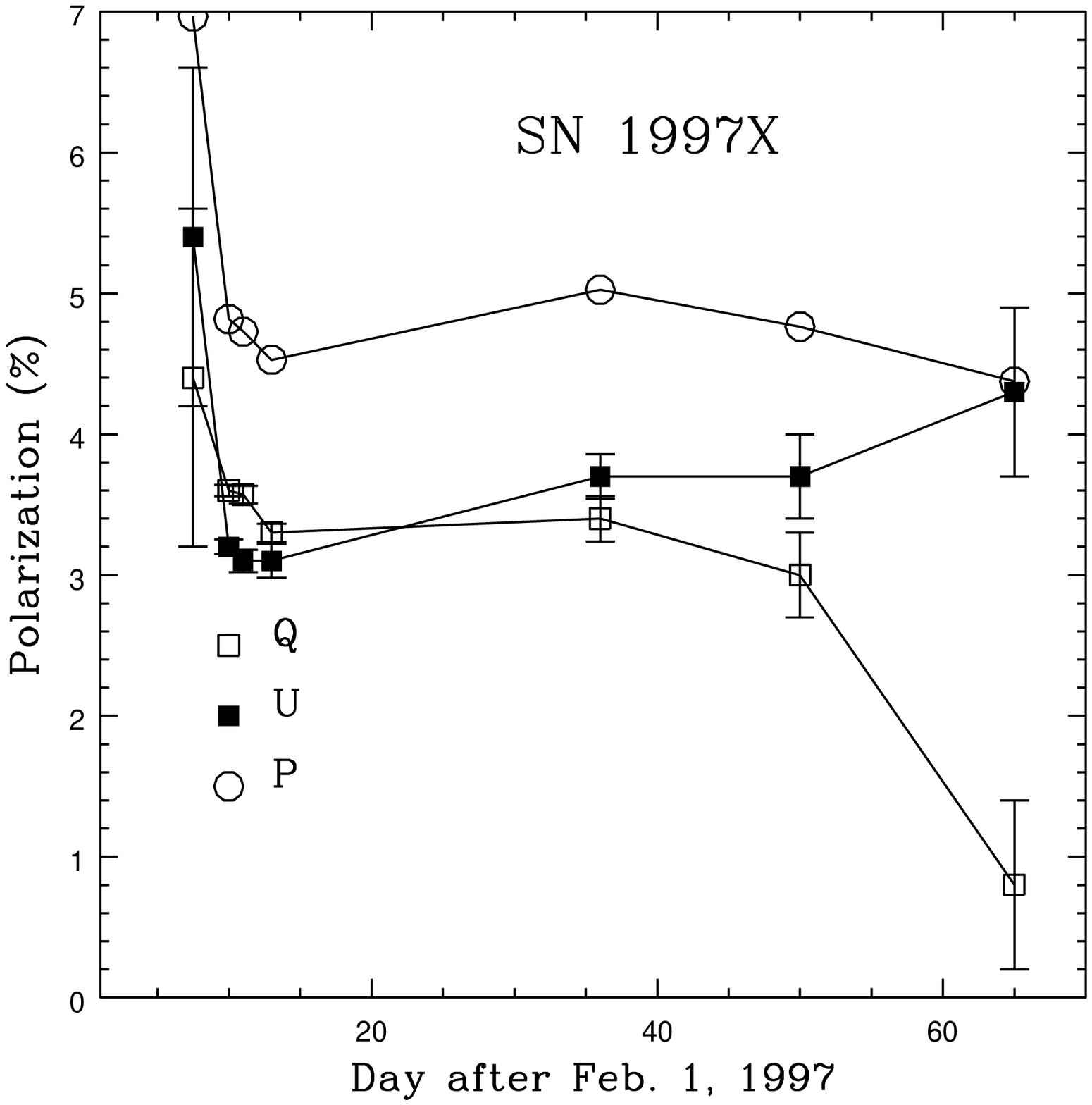}\includegraphics[width=4.cm,angle=0,height=4.5cm]{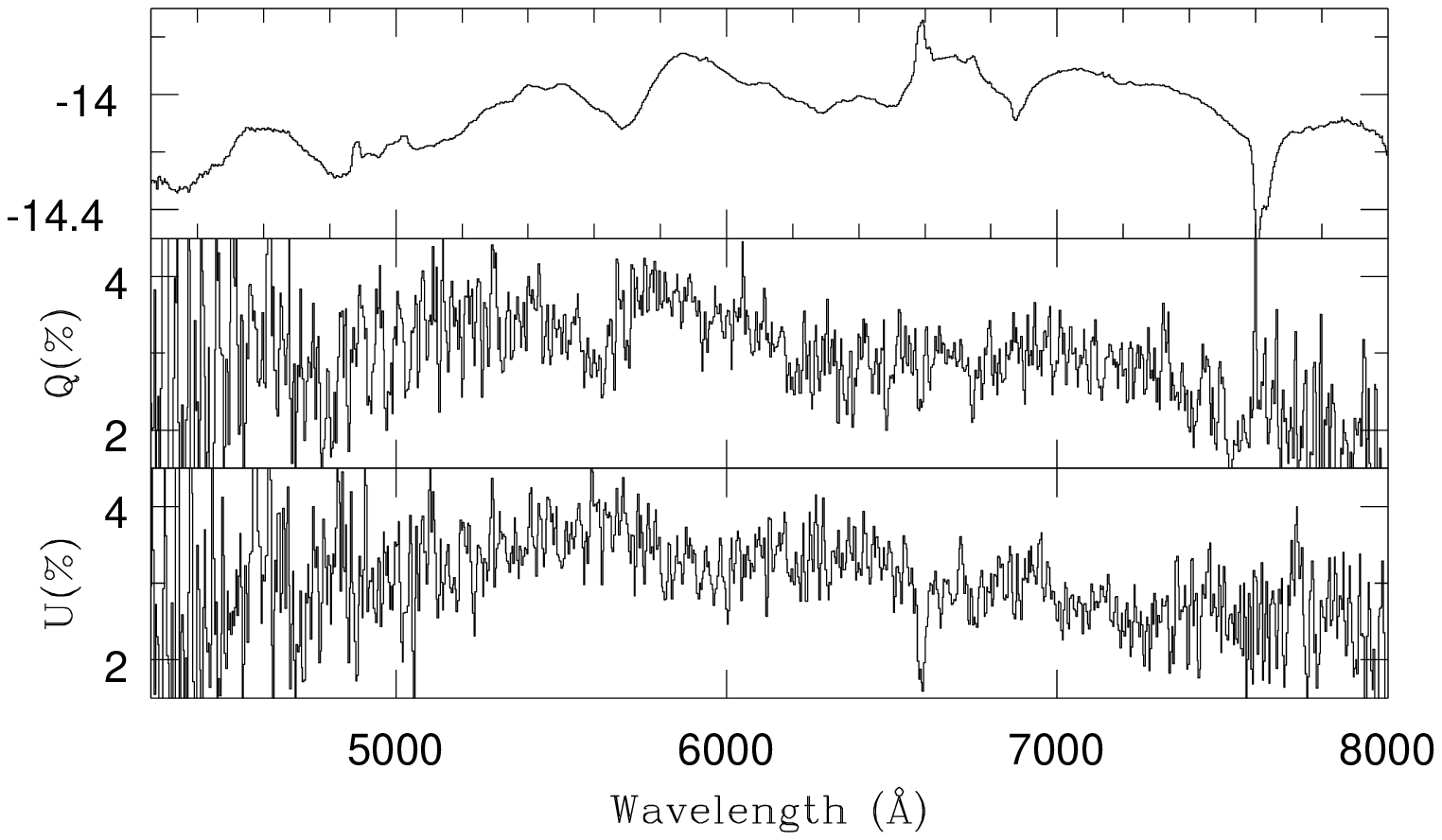}
\caption{Left: Evolution of the continuum polarization synthesized from the spectropolarimetry data. 
Right: Flux spectrum of SB 1997X on logarithmic scale and the observed degree of 
polarization obtained on Feb. 10, 1997 at the McDonald Observatory. Data at other epochs have similar quality.
The degree of polarization evolves with time. Conspicuous spectral features are detected in both Q and U.}
\end{figure}

SN 1997X is a typical Type Ic SN with no hydrogen or helium feature. This is an especially
exciting observation because SN~1997X, like all SN Ib/c, must have occurred in a nearly naked, non-degenerate,
carbon/oxygen core. There is little mass to dilute an asymmetrical explosion or to drive Rayleigh-Taylor
instabilities, and with rapid expansion, little time to propagate transverse shocks. There is the tantalizing
expectation that in SN~1997X we may be seeing the direct effects of asymmetric explosion (Wang et al. 2001). 
The degree of intrinsic polarization is the highest among all SNe with polarimetry observations which points to a disk
like geometry or a prolate spheroid with large major to minor axis ratios (see Fig. 1.3).
Models of core-collapse that, while convective, are spherical in the mean may be inadequate to induce 
the observed polarization unless instabilities at large scale can grow 
substantially (Blondin, Mezzacappa, \& DeMarino 2003). 
The large polarization of SN~1997X may mean that the effects of rotation and perhaps magnetic fields have to 
be included {\it ab initio} in the collapse calculations.

\subsection{SN 1998S}

SN~1998S is a Type IIn SN showing strong narrow H$\alpha$ emission lines. We have secured two
epochs of polarimetry data with the 2.1 meter telescope at the McDonald Observatory (Fig. 1.4; 
Wang et al. 2001) but more data are reported by Leonard et al (2000). Emission of Type IIn arises 
from ejecta-CSM interaction. SN~1998S showed a very well defined symmetry axis which can be associated 
with the geometry of the CSM matter around the SN. The degree of polarization approaches 3\%. In the 
extreme case that the polarization is modeled in terms of a central-source that is scattered by a CSM disk,
the opening angle of the disk would be on the order of a few degrees 
(McDavid 2001; Melgarejo et al. 2001; See also Wang et al. 2004 and the discussions on SN 2002ic).

\begin{figure}[hbt]
\includegraphics[width=10cm,angle=0]{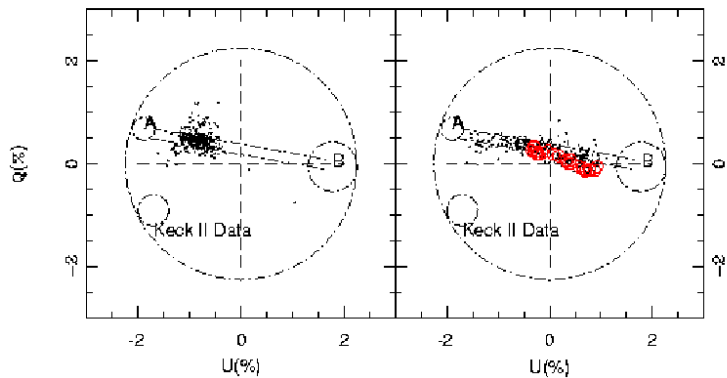}
\caption{Spectropolarimetry of the Type II plateau SN 1998S on the Q-U plane. The dot-dashed circles 
represent the limit of polarization caused by interstellar dust. Point A gives the actual location of the
interstellar polarization. The data points at the later epoch are dispersed along a well defined line
which indicates that the degree of polarization increased significantly during the two epochs of
observations.}
\end{figure}

\subsection{SN 1999em}

SN~1999em is a well observed Type II plateau SN. The polarization data show an extremely well defined
symmetry axis (Fig. 1.5). The degree of polarization increased sharply from before optical 
maximum to the plateau phase.

\begin{figure}
\includegraphics[width=8.cm,angle=0]{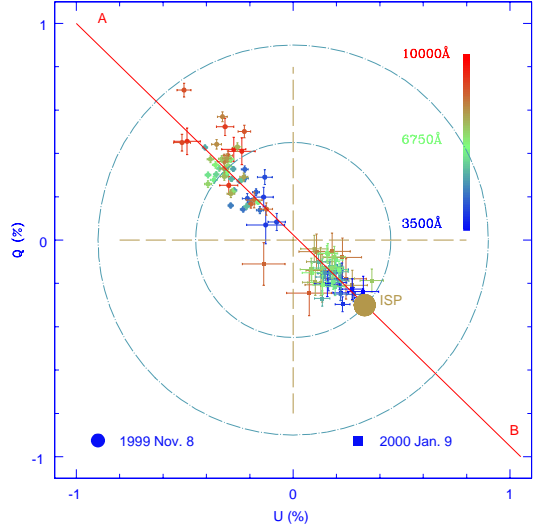}
\caption{Spectropolarimetry of the Type II plateau SN 1999em. The pre-max data cluster at the lower-right
quadrant, while the data at plateau phase are dispersed along a very well defined line - an indication of
a very well defined geometry with no significant small scale fluctuations.}
\end{figure}

\subsection{Hypernovae}

The association of peculiar Type Ic SN~1998bw with a faint GRB has led to the suggestion 
that SN Ib/c maybe responsible for some GRBs. The discovery of SN 2003dh/GRB 030329 proves these suggestions.
(Stanek et al. 2003; Matheson, 2003, these proceedings).
Spectropolarimetry are reported for SN~1998bw (Patat et al. 2001), and SN~2003dh (Kawabata et al. 2003).

Two other SNe with similar spectral properties were observed in polarimetry. SN~1997ef was
observed at the McDonald Observatory and the data shows that the degree of polarization can not
be significantly larger than 0.5\% (see Fig. 1.6), which is consistent with what were found for 
SN~1998bw and SN~2003dh.

\begin{figure}
\includegraphics[width=9cm,angle=0]{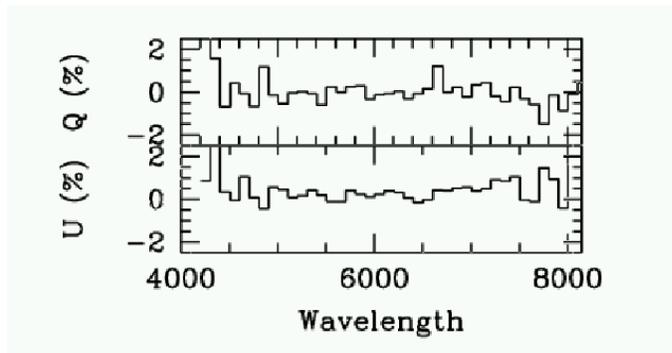}
\caption{Spectropolarimetry of the peculiar Type Ic SN~1997ef. The level of intrinsic polarization is $\le 0.5\%$}
\end{figure}

The nearby SN~2002ap is of similar spectral behavior but with no associated GRB. Extensive 
spectropolarimetry are obtained at the VLT (Wang et al. 2003b), Keck (Leonard et al. 2002), 
and the Subaru (Kawabata et al. 2002). In contrast to the low levels of observed polarization for 
SN 2003dh/GRB 030329, SN 1998bw and SN~1997ef, SN~2002ap was found to be rather highly polarized. 
The earliest data were from the ESO VLT and highly polarized features
from OI 7773 and Ca II IR triplet are detected (Fig. 1.7). The polarized features show sharp evolution 
with time, with the dominant axis of asymmetry changing with time. Post maximum observations 
show increases of the degree of polarization. These behaviors suggest that the ejecta are 
highly distorted at the central region, and thus point to a highly aspherical central 
engine for the explosion. These behaviors could be understood if the ejecta are ``jet-like'' with the jet pointing 
away from the observer for SN 2002ap, but head-on for SN 2003dh 
and SN~1998bw, making the projected geometry more symmetrical for the SN Ib/c 
with GRB counter-parts than those wihout. 

\begin{figure}
\includegraphics[width=12.cm, angle=0]{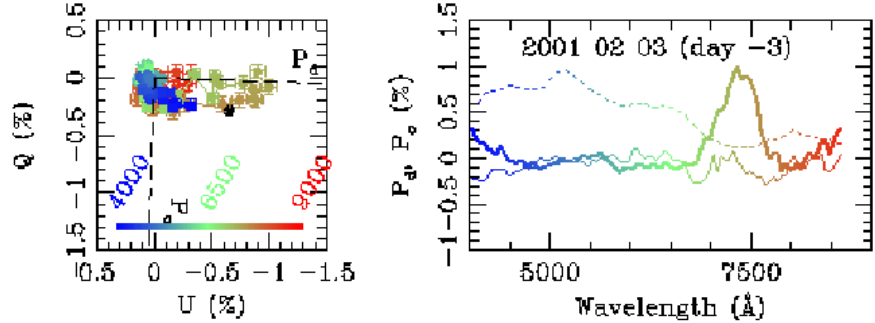}
\caption{Spectropolarimetry of the peculiar Type Ic SN~2002ap obtained on 
Feb. 03, about -3 days from $B$ maximum (the date of maximum is take from Foley et al. 2003). 
The data on the Q-U plane is shown on the left with the principle axis and the axis
orthogonal to it (the secondary axis) drawn in dashed lines. 
The flux, and the Stokes parameters projected onto the principle and secondary
axes are shown on the right hand side. The largest asymmetry 
is seen in the OI 7773 line and the Ca II IR triplet (Wang et. 2003b).}
\end{figure}
\section{Thermonuclear Supernovae}

Broad band polarimetry study showed that Type Ia SNe are in general less polarized than 
core-collapse SNe, with typical degree of polarizations less than 0.3\% (Wang et al. 1996). 
The polarization is thus hard to detect. SN~1992A was the only SN~Ia with spectropolarimetry 
observations before our program. Spectropolarimetry of SN~1992A two and seven weeks past
optical maximum were reported by Spyromilio \& Bailey (1992) and no significant polarization was 
detected. 

\subsection{SN 1996X}
SN~1996X was observed close to optical maximum, 
and the data are suggestive of intrinsic polarization at a level of $\sim$ 0.3\% (Fig. 1.8). 
It is realized that the interpretations of SN Ia polarization can be difficult due to
strong wavelength modulations of opacity. An asphericity of
11\% is derived from SN~1996X (Wang, Wheeler, H\"oflich 1997).

\begin{figure}
\includegraphics[width=9.cm,angle=0]{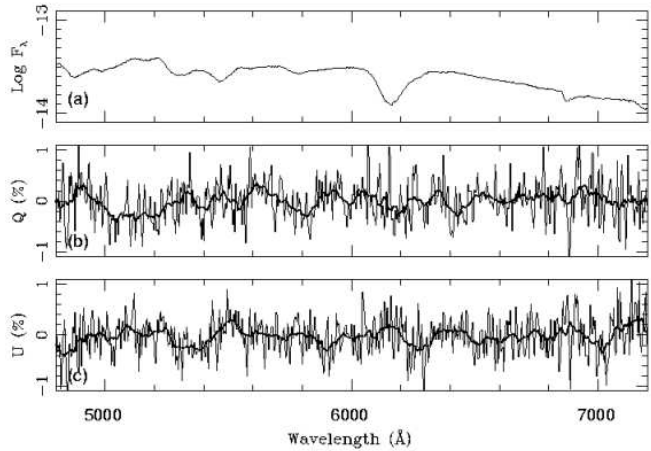}
\vspace{0.cm}
\caption{Flux spectrum (a), Stokes Q (b), and U (c) of the normal Type Ia SN 1996X obtained at
optical maximum. A low level of polarization of about 0.3\% is tentatively detected.}
\end{figure}

\subsection{SN 1999by}
SN 1999by is the first Type Ia with clear evidence of asphericity (Howell et al. 2001; Fig. 1.9). 
The degree of polarization is 0.7\% - comparable to those of core-collapse SNe. However, 
SN~1997cy is a subluminous Type Ia and the observed polarization does not probe the chemical 
layers in thermonuclear equilibrium. Nonetheless, it suggests that some SN~Ia may be aspherical 
despite the rather low level of polarizations found for other spectroscopically normal SN~Ia.

\begin{figure}
\includegraphics[width=6cm,height=6cm,angle=0]{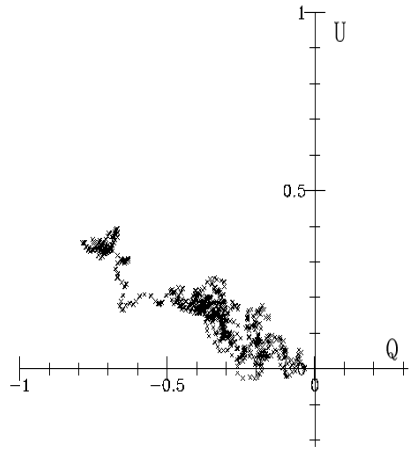}
\caption{Spectropolarimetry of the subluminous Type Ia SN 1999by showing that it is significantly
polarized, and thus highly aspherical. A well defined symmetry axis is observed. 
Note that the data points are smoothed and the errors are highly correlated. }
\end{figure}

\subsection{SN 2001el}
SN 2001el is the first {\it normal} supernova with multiple epoch observations covering first two months
after explosion (Wang et al. 2003a; Fig. 1.10). Thanks to the spectacular performance of the VLT, for the first 
time we obtained high quality data which can provide a clear picture of the geometrical structure of 
an SN Ia at different chemical layers. At the outermost region, a shell, a ring, or clumps of 
high velocity ($\sim 20,000-26,000$ km/sec) material, perhaps enriched in calcium is identified. This 
high velocity matter was found to be causing an absorption feature at 8000\AA\ 
and makes that feature polarized at 0.7\%. At around 10,000 km/sec, the Ca II and Si II lines show
polarizations around 0.3\%. Deeper inside, the materials are found to be spherical as evidenced
by a sharp decrease of the polarization degree past optical maximum. This indicates that there is no
significant chemical mixing in regions that were burned to thermonuclear equilibrium. 
Such a structure is consistent with what one expects from delayed detonation models but is in 
conflict with pure deflagration models. The observed degree of polarization suggests that the photosphere
can be aspherical at 10\% level and lead to magnitude dispersions of $\sim$ 0.1-0.2 mag.


\begin{figure}
\includegraphics[width=6cm,height=6cm,angle=0]{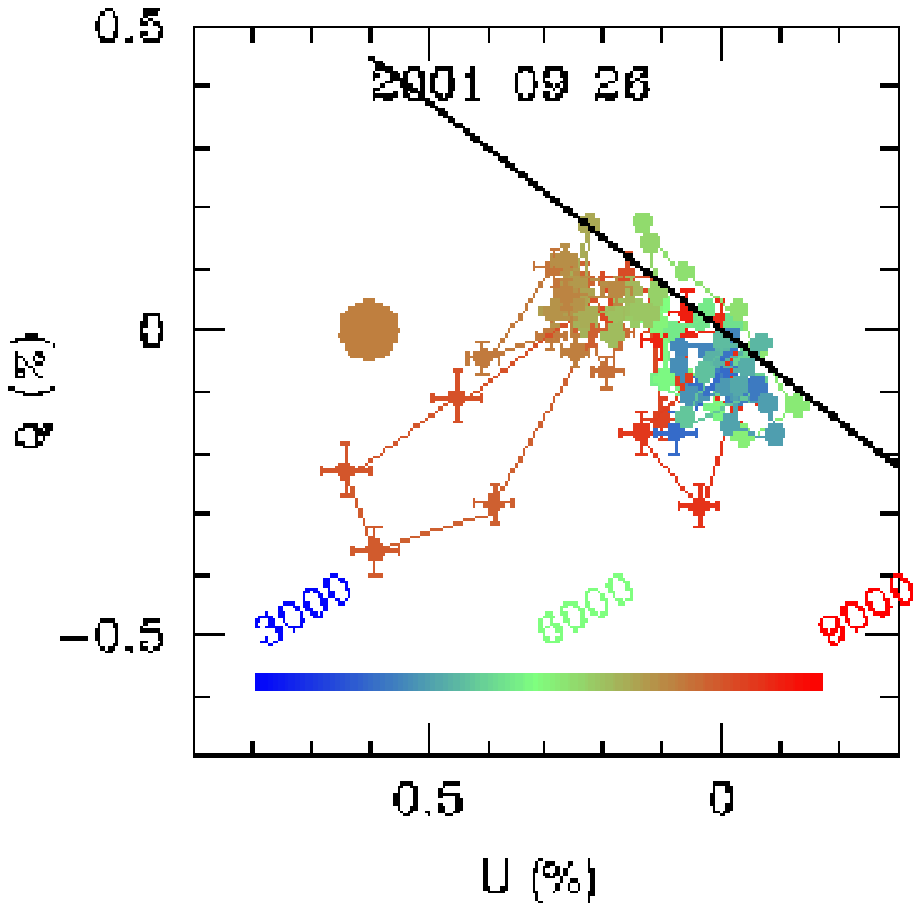}~\includegraphics[width=6.5cm,height=6.5cm,angle=0]{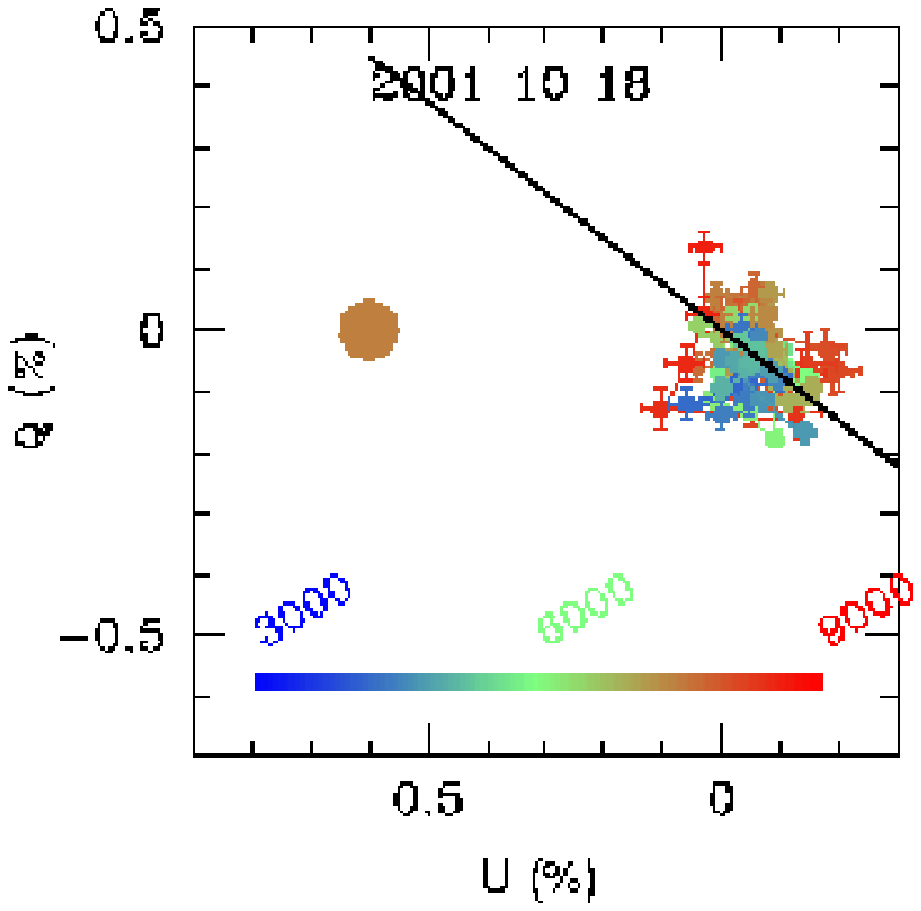}
\vspace{0cm}
\caption{Spectropolarimetry of the normal Type Ia SN 2001el. The polarization -4 days from optical maximum is
shown on the left hand side. The distribution of the data points on the Q-U plane indicates intrinsic
polarizations of about 0.3\% for the photosphere and 0.7\% for the high velocity Ca II line. About two weeks
past optical maximum, the polarization decreased substantially and became consistent with observational 
noise as shown on the right hand side panel.}
\end{figure}

\subsection{SN 2002ic}

SN~2002ic is the first SN Ia showing evidences of ejecta-CSM interaction (Hamuy et al. 2003). We have
obtained late time spectropolarimetry data at Subaru and the ESO VLT. The
required hydrogen mass is on the order of several solar masses with densities higher than 
10$^8$/cm$^{3}$. Such dense CSM must have small volume filling factor. The H$\alpha$ line
shows a depolarization of nearly 1\%. If the polarization is caused by scattering of
the supernova light by CSM matter, the CSM must be distributed in a dense
CSM disk (Wang et al. 2004; McDavid 2001; Melgarejo et al. 2001). The CSM matter bears 
striking resemblance to some well observed proto-planetary nebulae. This suggests that 
thermonuclear explosions may occur at all stages of the formation of a white-dwarf, with 
most of them, of course, at stages when the systems have no significant amount of CSM.

\begin{figure}
\includegraphics[width=9.cm,height=9cm,angle=0]{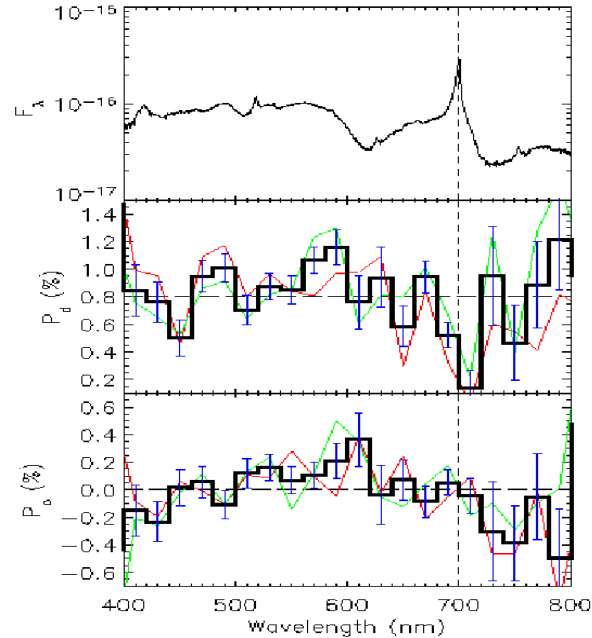}
\vspace{0.cm}
\caption{Flux (top), and the Stokes parameter projected onto the principle axis of asymmetry (middle)
and the axis orthogonal to the principle axis (lower) of the hybrid SN 2002ic. The Q and U for 
two different epochs are shown in thin lines and their weighted average are shown as thick histograms}
\end{figure}

\section{Summary}

Spectropolarimetry is still a rapidly growing field. It is the best method to probe the
geometrical structures of SNe. 

Our studies have shown that systematic polarimetry program of supernovae can provide 
detailed maps of the geometrical structures of the various chemical layers of 
supernova ejecta. These studies are important ingredients in the studies of the nature
of the supernova phenomena, and they are crucial in making SN~Ia better calibrated
standard candles.

The intrinsic luminosity dispersion introduced by the observed asphericity of SN~Ia is 
around 0.1-0.2 magnitudes which is of the same order of the residual luminosity dispersion 
of SN Ia after corrections of light curve shapes. The fact that SN Ia are aspherical 
before and at optical maximum implies that geometrical orientation along can introduce 
a significant amount of the luminosity dispersion. Since the luminosity-light curve shape
relation corrects mostly the effect of the amount of $^{56}$Ni, and is insensitive to the 
geometrical effect, it is then clear that SN Ia are indeed an extremely homogeneous 
group of events in their own right. As long as there is no significant evolution of the
geometrical structure of SN~Ia with redshift, geometrical effects can be corrected with a large enough
data sample; this is extremely simple compared to all the other possible complications 
such as progenitor chemical or mass evolution. Furthermore, SN~2001el also indicates 
that geometrical effect is much less 
prominent after optical maximum which makes these epochs extremely useful for deriving 
distance scales. It also makes the CMAGIC approach (Wang et al. 2003c) attractive as 
CMAGIC exploits exactly these later stages at which there is no substantial departures from spherical symmetry.

{\it Acknowledgment} I am grateful to D. Baade, P. H\"oflich, and especially J. C. Wheeler for 
collaborations on the polarimetry program. 
This work is supported by the
Director, Office of Science, Office of High Energy and Nuclear Physics, of the U. S. Department of
Energy under Contract No. DE-AC03-76SF000098. This work is partially based
on observations collected at the European Southern Observatory, Chile.

\end{document}